\documentclass[runningheads]{llncs}
\usepackage[T1]{fontenc}
\usepackage{graphicx}
\usepackage{url}
\begin{document}
\title{Equanimity in HRI: Applying Calm Technology Principles to Human-Robot Interaction}
\titlerunning{Equanimity in HRI}
%
\author{Barbara Sienkiewicz\inst{1}\orcidID{0009-0008-1977-5806} \and
Bipin Indurkhya\inst{1}\orcidID{0000-0002-3798-9209}}
\authorrunning{B.Sienkiewicz, B.Indurkhya}
%
\institute{Cognitive Science Department,
        Jagiellonian University, Kraków, Poland
\email{barbara.sienkiewicz@student.uj.edu.pl}\\}

\maketitle              
\begin{abstract}
This paper explores how {\textit{Calm Technology}} can be integrated into Human-Robot Interaction (HRI), with a particular focus on the household environment. It offers comprehensive guidelines for designing assistive robots that prioritize and enhance the human need for {\textit{equanimity}}, ensuring interactions are calm, non-intrusive, and harmonious. The paper examines the widespread influence of technology in contemporary life and its impact on cognitive capabilities, underscoring the need for responsible robotics and ethical considerations in future technological developments. By adapting {\textit{Calm Technology}} principles to domestic robots, the article provides concrete examples and features that should be employed in household assistive robotics. The goal is to foster a balanced, unobtrusive interaction between humans and robots, especially in the home environment, as it is the most privat environment in everyone's life, paving the way for applications and further research in the field.

\end{abstract}
\section{Introduction}

Four decades ago, a single computer often served multiple users. Today, the paradigm has changed dramatically, with numerous computing devices serving a single individual \cite{case2015calm}. It is common to possess a personal computer, a smartphone; and increasingly devices such as smartwatches and cars are equipped with built-in computers. This number of technologies, while offering unprecedented convenience and connectivity, can also lead to a sense of being overwhelmed.

In this situation, the idea of {\textit{Calm Technology} comes up as a promising oasis. First introduced by Weiser and Brown in 1995 \cite{weiser1996designing}, {\textit{Calm Technology} focuses on designing technology that works in the background, demands minimal user attention and allows focus on other tasks without being overwhelmed by constant notifications, alerts, or interactions. The essence lies in its ability to engage both the central and peripheral aspects of human attention, smoothly shifting between them as needed \cite{weiser1996designing}.

Despite its inception nearly thirty years ago,{\textit{Calm Technology} principles have not influenced mainstream technology design as much as one might expect. We advocate here a renewed emphasis on {\textit{Calm Technology}, recognizing its relevance especially in the current era, where the intrusion of digital devices into every aspect of our lives requires a paradigm shift toward more calm and less intrusive technologies.

This work concentrates on the application of {\textit{Calm Technology} principles within Human-Robot Interaction (HRI), with a focus on the household environment by providing guidelines for the design of assistive robots.
\section{Psychological impact of overstimulating technology}
In our modern lives, the presence of technology often crosses the threshold from being a tool for convenience, to a source of constant stimulation. 
The increasing number of devices that require our attention can sometimes (or often) lead to overstimulation. This occurs when the brain attempts to process too much information, leading to a decline in cognitive function \cite{kershaw2023extent}, which can result in mental health problems such as depression, anxiety, and attention deficit disorder \cite{schultz2023technology,bucci2019digital,small2020brain}.
The phenomenon of being overwhelmed by technology, particularly within healthcare systems, has been extensively studied \cite{sendelbach2013alarm,bridi2014clinical}. Medical professionals, including nurses and doctors, often encounter a barrage of auditory signals from various medical devices. This constant exposure to alarm sounds can lead to desensitization, where the ability to respond appropriately to these warnings diminishes over time. This condition, referred to as 'alarm fatigue', leads to ignoring critical alerts or responding inadequately to them \cite{cash2009alert}. This situation underscores the need for careful consideration of the psychological impact of technology in all environments, not limited to healthcare. Overstimulation with digital technology, which induces frequent dopamine releases, can negatively affect motivation and well-being. Implementing digital detox strategies can mitigate these effects and help restore a healthier psychological state. However, technology should not be designed so that detox is required.
'Digital detox', defined by the Oxford Dictionary as a period during which a person refrains from using digital or electronic devices to break a pattern of excessive or compulsive use and prioritize face-to-face social interactions, mindfulness, and harmony with nature \cite{digital_detox}, highlights the imbalance in our relationship with technology. This trend also reflects an increasing awareness of the psychological impact of technology on individuals. 

It is important to recognize that not all people feel overwhelmed to the same degree in response to external stimuli; individual differences and coping mechanisms play a significant role in the moderating of these experiences \cite{schlossbereg1999overwhelmed,venkatesh2000longitudinal}. However, there are inherent limitations to human cognitive and attentional capacities that affect everyone. Therefore, it is crucial that technological systems are designed in a manner that does not overly tax or extend these capacities, thereby avoiding reliance on users' ability to stretch their cognitive resources beyond sustainable limits.

\section{Responsible robotics}
As Peter-Paul Verbeek, a contemporary philosopher of technology, points out, technologies significantly shape the relationship between humans and their world, offering diverse and enriched encounters with that world \cite{verbeek2005things}. This shaping can be seen as both an opportunity and a danger, depending on how these technologies are designed. Therefore, it is crucial to focus on responsible robotics, which requires considering the ethical implications of technology before it is fully developed, as described in \cite{urquhart2019responsible}. This approach ensures that designers, engineers, and scientists are aware of these ethical issues as they develop the technology \cite{coeckelbergh2022robot}. It is a collective responsibility, as different stages of development may introduce different ethical and usability concerns.
The focus on human well-being and mental states is crucial, advocating for a holistic view that goes beyond mere functionality to consider technology's broader impact on society and the environment. This approach, aligned with the principles of {\textit{Calm Technology}, emphasizes the design of technologies that improve user well-being and environmental harmony - {\textit{equanimity}} -  aiming to create a cohesive ecosystem that benefits all its agents.


Understanding Calm Technology within the market requires a shift in how both consumers and companies perceive and interact with technology. Instead of competition, companies should aim for synergy, creating ecosystems where products complement each other without overwhelming users. Prioritizing consumer needs and focusing on quality and integration over quantity can lead to a technology landscape that enriches, rather than clutters, our lives. However, implementing this approach is challenging given the current market behavior. In addition, when properly designed, robots can help maintain sustainable behaviors, facilitating social inclusion or being a companion, as discussed in \cite{indurkhya2024robots}.

\section{Application of Calm Technology principles to HRI}

In this section, we explore the {\textit{Calm Technology} principles \cite{case2015calm} can be incorporated into HRI, focusing on their importance for current and future home environments. 

\subsection{Technology should require the smallest possible amount of attention}

\begin{quote}
    `The more things you have to pay attention to, the less mental space you have available for actually getting things done, and the more stressful those interactions are going to be.' \cite[p.~32]{case2015calm}. 
\end{quote}

This concept is beautifully illustrated by the inefficiency of multitasking. What is often perceived as multitasking, the ability to engage in several activities at once, is actually the ability to quickly switch tasks \cite{salvucci2009toward}. Technology should deliver messages without interrupting your current task, much like a waiter serving your coffee without interrupting your conversation.

Based on the idea of {\textit{Calm Technology}, robots should aim to integrate seamlessly into the user's environment, providing assistance and information without being obtrusive. This can be achieved through ambient notifications, subtle visual cues, or background processes that surface only when user intervention is required or when attention is already directed to the robot.

\subsection{Technology should inform and create a sense of calm}

\begin{quote}
    `The role of technology in creating a sense of calm is pivotal, particularly through its ability to communicate effectively and reassure users of its proper functioning.' \cite[p.~32]{case2015calm}.
\end{quote}
Technology can convey calmness by providing clear indications that the systems are operating correctly. The research of Tatsuya Nomura \cite{nomura2011relationships} suggest that when robots share self-disclosure information, it helps reduce human anxiety about communicating with them. For example, subtle visual cues, such as a light on a robot's head indicating that it is {\em thinking,}, can inform users that the robot is processing information but not yet ready to respond. This gentle form of communication reassures users that everything is under control without causing unnecessary anxiety or interruptions. This is related to the concept of {\textit{transparency}} in robotics: the degree to which the user understands the ability, intention and situational constraints of the robot \cite{wortham2017robot}. Rossi and Rossi \cite{rossi2024way} emphasise that transparency in human-robot interaction is significantly improved through real-time explanations of robot actions. They point out that transparency is essential not only in interactive tasks but also in noninteractive tasks, as providing clear indications of the robot's status and intentions helps to sustain user trust.

Effective technology design also involves an ability to alert users during critical events. Systems should employ multiple sensory signals: auditory beeps, visual alerts, and even haptic feedback — to ensure that important notifications are noticed \cite{haas2014multimodal}. For instance, smoke detectors use both sound and light to grab attention during emergencies, or an alert indicating a critical issue, such as a potential gas leak or electrical fault, must be immediate and non-overrideable to ensure user safety. However, notifications about less critical issues, such as reminders to turn off lights or notifications about high energy consumption during peak hours, can be personalized. Users can choose to receive these notifications based on their personal energy-saving goals and daily routines, allowing the system to support their lifestyle without causing unnecessary stress or interruptions.

\subsection{Technology should make use of the periphery}
\begin{quote}
    `A calm technology engages both the center and the periphery of our attention, and in fact moves back and forth between the two.' \cite[p.~2]{weiser1996designing}
\end{quote}
This principle is related to the first one; however, it emphasizes the role of peripheral aspects of human perception. The peripheral aspects of human perception can be effectively engaged by smoothly transitioning attention from the center to the periphery and back or by enhancing peripheral details \cite{tugui2004calm}. This approach suggests that information delivery does not always require direct or focused attention. Instead, information can be conveyed passively, without demanding additional actions for verification. For example, a robot can clearly indicate its intended path to prevent human users from worrying about obstructing its way. This can be accomplished through visual indicators such as lights, projected paths on the floor, or the robot's gaze direction, as discussed by Hart et al. (2020) \cite{hart2020using}. Such methods ensure that users are passively informed of the robot’s movements without the need for active monitoring.

\subsection{Technology should amplify the best of technology and the best of humanity}
\begin{quote}
    `Machines shouldn’t act like humans at least, not in the current design environment that lacks the framework to effectively integrate “humanness” with devices — and humans shouldn’t act like machines.' \cite[p.~32]{case2015calm}
\end{quote}
Amber Case expresses in this principle the distinct, non-substitutable roles of humans and technology. She argues that the essence of human experience is creativity, emotional depth, learning, and the ability to make mistakes. The realm of technology, on the other hand, is defined by computational precision, consistency, and reliability \cite{case2015calm}. While these roles may appear contradictory, this contrast does not necessarily lead to stress. Instead, it is essential to ensure that technology enhances human autonomy and emotional well-being without imposing judgment or discomfort. It involves giving people autonomy in decision-making, free from technological judgment or guilt over their emotions. To implement this principle effectively in a home environment, it is important to ensure that technology supports human autonomy and emotional well-being without encroaching on personal space or evoking a feeling of being judged by technology. Although feeling watched may not necessarily violate one's sense of humanness, it can significantly affect an individual's comfort and the ability to express himself freely. Studies indicate that even small cues can make people feel like they are being watched \cite{van2023reputation,hesslinger2017sense}, and robots also awaken these feelings, as shown in \cite{9802975,shinohara2023understanding}. However, there remains a gap in understanding how these sensations manifest when the individual is focused on a different task, which should be researched to fulfill the 4th principle.

\subsection{Technology can communicate but doesn't need to speak}

\begin{quote}
    `A user interface that requires all our visual focus distracts us from doing anything else. An interface that requires our complete auditory focus (or perfect enunciation) is equally distracting.' \cite[p.~36]{case2015calm}
\end{quote}

Voice interfaces are convenient, but often lack reliability in their operation \cite{clark2019state}. As an alternative to speech, technologies can employ non-verbal auditory cues to convey messages effectively. For instance, a short, cheerful melody can signal the completion of a task by a robot, while a distinct, sad tone could indicate the need for assistance. This approach extends beyond the auditory channel, enabling auditory signals to communicate statuses or needs universally without language barriers.

In addition, audio is not the only channel for communication; visual information is always an option. For example, visual cues, such as lights or on-screen notifications, can provide a meaningful context. In addition, haptic feedback, such as vibrations or tactile sensations, offers another channel of communication.

Integrating verbal auditory cues with gesture-based interactions can enhance the naturalness of human-robot interactions. This integration not only makes the interaction more intuitive for humans, but also provides an additional source of information when auditory cues are interrupted or misleading \cite{wu2021multimodal}.

This principle highlights the potential of technology to utilize alternative communication methods, offering intuitive and accessible interactions that do not rely on verbal communication.

\subsection{Technology should work even when it fails}

\begin{quote}
    `The edge cases are where things go wrong (...). But the fact is that everyone is an edge case at one time or another.' \cite[p.~39]{case2015calm}
\end{quote}
In the use of robots, there will always be edge cases where the robot might not behave as expected. Simply, it is crucial that users have the ability to turn off the robot if it malfunctions or does not perform its tasks correctly, or the user does not want to use it anymore. Providing a simple and reliable shutdown option ensures user safety and confidence, allowing better robot management in any situation, including unexpected or unusual conditions. 
Typically, robots are equipped with such emergency shutdown buttons. For example, the robot Pepper has this button on its back just below the neck \cite{miyagawa2019consideration}, while the robot Tiago has it located on its base platform \cite{pages2016tiago}. 

\subsection{The right amount of technology is the minimum needed to solve the problem
}

\begin{quote}
    `You shouldn't have to be a system administrator to live in your own home. And you shouldn’t have to have a system administrator.' \cite[p.~44]{case2015calm}
\end{quote}
Paraphrasing Amber Case to apply to robotics for home environments: You shouldn't have to be a roboticist to use a robot in your home, and you do not need a roboticist to do so.
This statement underscores the importance of user-friendly design in home robotics, emphasizing simplicity, reliability, and robustness. The design of robots should prioritize the user experience, ensuring that anyone can operate the robot without extensive training or technical support. This includes designing both hardware and software to meet the needs and behaviors of the average user. For instance, error messages should be easily understandable and provide clear instructions for resolution.

Furthermore, the initial setup process for home robots should be minimal and straightforward. Ideally, this would involve plug-and-play systems where the robot is ready to use right out of the box, with minimal assembly or configuration required. Any necessary setup steps should be clearly outlined and easy to follow.

\subsection{Technology should respect social norms}

\begin{quote}
    `A society’s cultural norms define the social forces that push humans to interact in a way that is congruent with accepted social rules.' \cite[p.~44]{case2015calm}
\end{quote}
We argue that the same rule should apply to robots that interact with humans. In addition, a robot’s goals should be achieved by following these social norms, for example, a robot who wants to enter an elevator with people already inside should first wait for passengers to exit the elevator \cite{rossi2020secret}.  

Moreover, there should be a standardized ontology of robots that is known by the user \cite{haidegger2013applied}.
However, it is important to note that social rules and ontologies are not universally applicable. Research has shown that cultural differences significantly impact HRI \cite{lim2021social,ornelas2023redefining,rau2009effects}, which presents a salient challenge in designing a suitable robot behavior.

This challenge emphasizes the need for a new, user-centered design approach. Studies have extensively criticized the old approach and discussed the importance of adapting design processes to account for diverse cultural contexts \cite{rosner2018critical,shew2023against}. Consequently, roboticists must shift their focus to creating designs that are more inclusive and mindful of the varying social norms across different cultures. A participatory design approach, also known as co-design, is recommended where the target user group is included in the design process, and many local solutions are developed that address the specific needs of different user groups \cite{sienkiewicz2024participatory}.

Each of these eight principles of applying \textit{equanimity} to HRI is visually depicted in Figure \ref{fig1}.
\begin{figure} [h!]
\centering
\includegraphics[width=0.6\textwidth]{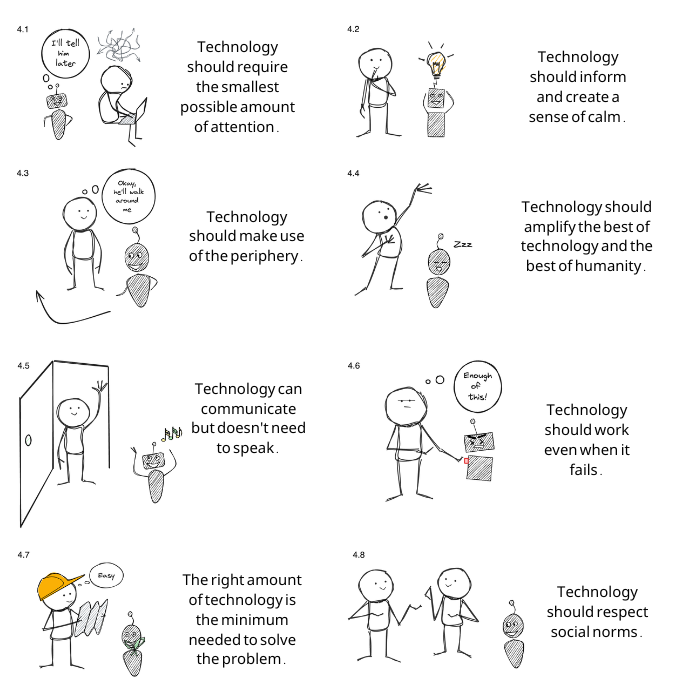}
\caption{Use cases of applying equanimity to HRI} 
\label{fig1}
\end{figure}

\section{Experimental Validation}

The \textit{Equanimity} concept while theoretically robust, requires rigorous experimental validation to establish its practical efficacy. Although longitudinal studies would offer more comprehensive insights, current technological limitations necessitate a focus on short-term experiments that can later be synthesized into broader long-term studies. Each principle can be independently validated through targeted experimental studies, forming the foundational steps toward developing unobtrusive domestic robots. The concrete examples for the individual validation of each principle are presented in first author master's thesis \cite{mythesis}.

Several existing studies already support the principles underlying Equanimity in HRI. For instance, a study by Smith et al. \cite{9980647} underscores the significance of balancing user autonomy with robotic assistance, which aligns with the first principle. This principle emphasizes the importance of empowering users while ensuring effective robotic support, a balance crucial for user acceptance. Similarly, the second principle, which pertains to transparency in HRI, advocates for the provision of clear and accessible information to users about the robot's actions. Although research in this area has predominantly focused on one-on-one interactions where users devote full attention to the robot in controlled laboratory settings, further investigation is needed to explore scenarios in which user attention is divided between the robot and various tasks. This approach more accurately reflects the real world.

In addition, the third principle involves effective communication of a robot’s intentions, such as its movement trajectory, which is critical not only for domestic settings but also for robots operating in public spaces. Preventing collisions without capturing the attention of bystanders is essential for seamless human-robot co-existence. The implementation of these principles can be evaluated using standardized metrics such as the Godspeed questionnaire \cite{bartneck2023godspeed}, which measures various dimensions of robot acceptance.

Given the inherent complexity of human behavior, a universal approach to robot design is inadequate. It is imperative for roboticists to engage with specific target groups from diverse cultural backgrounds to ensure that the robot's behavior is appropriately tailored. Recognizing cultural variability in human behavior, participatory design emerges as a valuable approach. For example, the participatory design framework detailed in \cite{sienkiewicz2024participatory} provides a robust basis for research aimed at understanding the needs and social norms of the elderly, thus facilitating the development of effective robotic systems for this demographic.

\section{Discussion}
Current HRI frameworks focus on the dynamics of interaction between a human and a robot, often neglecting the crucial influence of the environment or focusing on isolated aspects of the interaction, as demonstrated by Lyons \cite{lyons2013being} and Honig \cite{honig2018understanding}. Although these models offer valuable insights, they underscore the need for a more holistic approach. Therefore, we shift our perspective from dissecting individual elements of the interaction to considering the entire space in which these interactions occur, such as a household.
The integration of interactive robotics into our daily environments, especially within domestic settings, remains an underappreciated frontier. 

From a commercialization point of view, the lack of perceived necessity among consumers and businesses for domestic robots poses a significant barrier. Without clear market demand, there is insufficient momentum driving the development and adoption of such technologies. Additionally, designing robots that can seamlessly integrate into the household environment presents a complex and costly challenge, requiring innovations that go beyond the current technological capabilities of individual companies.

For researchers, the journey from laboratory experiments to real-world application is full of difficulties. Although initial studies conducted in controlled settings provide valuable preliminary data, the true test of these systems lies in long-term studies conducted within the actual homes of participants. This step is not only logistically challenging, but also ethically and practically complex, as it requires access to private domestic spaces and the sustained involvement of participants.

Furthermore, the successful development of such robotics systems will require the collaboration of a multidisciplinary team, including designers, psychologists, and other specialists, each bringing their unique expertise to bear on the project. Although these challenges may appear daunting, they should not deter us from pursuing this research. The potential implications of developing robots that are compatible with the human needs and the environmental constraints both can be extended far beyond households to include schools, hospitals, and public spaces \cite{niemela2021robots}.

\section{Conclusions}

This paper has explored the integration of {\textit{Calm Technology} principles within Human-Robot Interaction (HRI), with a particular focus on the design of household robots that facilitate calm, non-intrusive interactions. By prioritizing the concept of equanimity, this research offers a framework for developing robots that are mindful of human cognitive limits and aim to enhance quality of life without overburdening users. Although these guidelines provide a foundational approach, the practical implementation of such technologies is fraught with challenges.  However, the insights presented here take a significant step toward a deeper understanding of how robots can be effectively integrated into everyday life. Continued research will be essential to overcome practical, ethical and cultural challenges, ultimately paving the way for the successful deployment of these technologies.

\subsubsection{\ackname} 
Funded by the National Science Centre, Poland under the OPUS call in the Weave programme under project number 2021/43/I/ST6/02489.

\subsubsection{\discintname}
The authors have no competing interests to declare that are
relevant to the content of this article.
\bibliographystyle{IEEEtran}
\bibliography{export}

\end{document}